\documentclass[aps,prl,twocolumn,showpacs,superscriptaddress,groupedaddress]{revtex4}  

\usepackage{amsmath}
\usepackage{amssymb}
\usepackage{amsfonts}
\usepackage{float}
\usepackage{graphicx}
\usepackage{tabularx}
\usepackage{multirow}
\usepackage{lscape}
\usepackage{rotating}
\usepackage{pstricks}

\begin{document}

\title{Extended Skyrme pseudo-potential deduced from infinite matter properties}


\author{D. Davesne}
\affiliation{Universit\'e de Lyon, F-69003 Lyon, France; Universit\'e Lyon 1,
             43 Bd. du 11 Novembre 1918, F-69622 Villeurbanne cedex, France\\
             CNRS-IN2P3, UMR 5822, Institut de Physique Nucl{\'e}aire de Lyon}
             \author{J. Navarro}
\affiliation{IFIC (CSIC-Universidad de Valencia), Apartado Postal 22085, E-46.071-Valencia, Spain}
\author{P. Becker}
\affiliation{Universit\'e de Lyon, F-69003 Lyon, France; Universit\'e Lyon 1,
             43 Bd. du 11 Novembre 1918, F-69622 Villeurbanne cedex, France\\
             CNRS-IN2P3, UMR 5822, Institut de Physique Nucl{\'e}aire de Lyon}
             \author{R. Jodon}
\affiliation{Universit\'e de Lyon, F-69003 Lyon, France; Universit\'e Lyon 1,
             43 Bd. du 11 Novembre 1918, F-69622 Villeurbanne cedex, France\\
             CNRS-IN2P3, UMR 5822, Institut de Physique Nucl{\'e}aire de Lyon}
             \author{J. Meyer}
\affiliation{Universit\'e de Lyon, F-69003 Lyon, France; Universit\'e Lyon 1,
             43 Bd. du 11 Novembre 1918, F-69622 Villeurbanne cedex, France\\
             CNRS-IN2P3, UMR 5822, Institut de Physique Nucl{\'e}aire de Lyon}
\author{A. Pastore}
\affiliation{Institut d'Astronomie et d'Astrophysique, CP 226, Universit\'e Libre de Bruxelles, B-1050 Bruxelles, Belgium}


\begin{abstract}
We discuss the contributions to the Equation of State for the N$\ell$LO Skyrme pseudo-potential ($\ell$=2,3). We show that by adding 4th and 6th order gradient terms, it is possible to fairly reproduce the spin/isospin decomposition of an equation of state obtained from \emph{ab-initio} methods. Moreover, by inspecting the partial-wave decomposition of the equation of state, we show for the first time a possible way to add explicit constraints on the sign of the tensor terms of the Skyrme interaction.
\end{abstract}


 
\date{\today}


\maketitle

\noindent \emph{Introduction}. A very successful nuclear physics model adopted to describe nuclear properties from drip-line to drip-line is the Nuclear Energy Density Functional (NEDF) theory~\cite{ben03}. Recently NEDF has been also applied to study compact astrophysical objects as neutron stars~\cite{tod05,gor09,erl13}. To this purpose, the functional should be able to describe some infinite matter pseudo-observables. 
Among the different available functionals, the one derived from the effective non relativistic Skyrme interaction~\cite{sky59} is probably the most used by the nuclear structure community. In its \emph{standard} form~\cite{per04}, the Skyrme functional consists of a linear combination of local densities up to second order in gradients plus a density dependent  zero-range term which replaces the original three-body  term~\cite{vau72,sad13}. 
Due to its very simple structure, Brink and Vautherin~\cite{vau72} performed the first numerical spherical Hartree-Fock (HF) calculations of even-even nuclei  already in the early 70s.
Once the structure of the functional has been chosen, one should determine its coupling constants. This is done by building a penalty function and then performing a minimization in a multi-dimensional space~\cite{dob14}.
The scientific collaboration named  UNEDF-SciDAC~\cite{ber07,fur11} has investigated the role of the optimization procedure on the quality and predictive power of the \emph{standard} Skyrme functional~\cite{kor10}. In their last article~\cite{kor13}, they have shown that there is no more room to improve the spectroscopic qualities of the \emph{standard} Skyrme functional, by simply acting on the optimization procedure. Two possibilities are then available: (i) following the DFT theory, where the primary building block is the functional that includes all correlation effects, (ii)  following the self-consistent mean-field theory, where the major ingredient is an effective pseudo-potential and correlations are added afterwards using a beyond mean field approach.
The main advantage of the second way is that we can follow a precise hierarchy including explicitly beyond-mean-field~\cite{ben03} correlations towards the exact many-body ground state~\cite{yan07}.

Following the second approach, Lesinski \emph{et al.} have investigated the inclusion of a tensor interaction~\cite{les07}, while Sadoudi \emph{et al.} considered an explicit central three-body term~\cite{sad13}. Carlsson \emph{et al.}~\cite{car08} have studied the possibility of including higher order derivative terms into the Skyrme functional. By means of the Density Matrix Expansion method~\cite{car10}, they have shown that it is possible to expand a finite range interaction in terms of a zero range like and such an expansion converges. In particular comparing the resulting binding energies of some selected nuclei, they have shown that using the \emph{standard} Skyrme functional up to second order (N1LO), the average discrepancy is of the order of $\approx10$ MeV, while adding the 4th order terms (N2LO) the error decreases by one order of magnitude.
Despite these  encouraging results, the major problem posed by this kind of approach lies in the strategy adopted to determine the higher order parameters. A badly constrained coupling constant can lead  to numerical instabilities in finite nuclei calculations~\cite{les06}.
In Ref.~\cite{hel13}, a numerical criterion based on the linear response (LR) formalism~\cite{pas15}  in symmetric nuclear matter (SNM) has been presented to avoid this kind of pathologies.  In Refs.~\cite{bec14,dav14k}, we have given explicit formulas for the N$\ell$LO cases discussed here, and we have also derived the necessary equations in coordinate space to be used in a numerical code. We also recall that the numerical code HOSPHE~\cite{car10b} can already deal with higher order functionals (N3LO) to compute ground state energies of spherical even-even nuclei using the Harmonic Oscillator basis. 
Combining all these informations, it is thus possible to build an optimization procedure to determine the parameters of the functionals avoiding by construction all bulk instabilities~\cite{pas13}. 

In this letter, we present a novel strategy to properly  constrain order by order the different parameters of the extended pseudo-potential by comparing with \emph{ab-initio} calculations in infinite nuclear matter (INM).
The use of INM pseudo-observables to constrain the Skyrme functional has been already discussed by Chabanat \emph{et al.}~\cite{cha97}; in particular they included in the penalty function the numerical values of the Equation of State (EoS)  in pure neutron matter (PNM) of Wiringa~\emph{et al.}~\cite{wir95}. In recent years, we have seen a remarkable effort of the \emph{ab-initio} community   to improve the quality of their calculations in INM~\cite{bal97,tew13,vid09}. All these  data represent a valuable opportunity to constrain the coupling constants of the N$\ell$LO functional.
In fact, several \emph{ab-initio} methods give us not only the general EoS in SNM, but also in each spin/isospin channel, and in some cases the partial wave contributions. In Ref.~\cite{les06}, Lesinski \emph{et al.} have shown that a \emph{standard} Skyrme functional 
is not flexible enough to reproduce these extra pseudo-data, otherwise it would be over-constrained and it would be incompatible with  finite nuclei observables.
There is in fact a strong disagreement, as shown in Fig.~6 of Ref.~\cite{les06}, 
not only in magnitude, but also in the sign of the energy per particle in the different $(S,T)$ channels as a function of the density. 
In the present letter, we show that the higher order derivative terms lead to a fair agreement with these pseudo-data. Moreover, 
by using the partial waves decomposition of the EoS~\cite{dav14k}, it is possible to put constraints on the sign of the tensor and spin-orbit N$\ell$LO terms. Such constraints are very useful since at present, because the existing parameterizations, based on ground-state observables of atomic nuclei, span in fact over a wide range of magnitudes and signs~\cite{les07}.

\emph{Results}. 
The extended Skyrme pseudo-potential (N$\ell$LO) contains central, spin-orbit and tensor components, whose general expressions have been deduced in Refs.~\cite{bec14,dav14k}. Besides the zeroth-order parameters $t_0^{(0)}, x_0^{(0)}$, the parameters are labelled as  $t_1^{(n)}, x_1^{(n)}, t_2^{(n)}, x_2^{(n)}$ for the central part, and $t_o^{(n)}, t_e^{(n)}$ for the tensor part, where $n=2 \ell$ is the order of the N$\ell$LO expansion. With this notation, terms involving $t_0^{(0)}, t_1^{(n)}, t_e^{(n)}$ are even with respect to space exchange, and the remaining terms are odd. There is a single spin-orbit term, with coefficient $W_0$, which is originated from the second order expansion~\cite{dav13}.

The EoS for the N${\ell}$LO pseudo-potential  in SNM can be easily obtained by means of a simple HF calculation~\cite{dav13}. 
Using the coupled spin-isospin basis $(S,T)$ it is written as
\begin{equation}\label{eos:st1}
E/A= \frac{3}{5} \frac{\hbar^2}{2 m} k_F^2 +\sum_{S,T} {\cal V}^{(S,T)} \,,
\end{equation}
where $k_F=( 3 \pi^2 \rho/2)^{1/3}$, with $\rho$ being the density, and ${\cal V}^{(S,T)}$  is the potential energy per particle projected onto the different spin/isospin sub-spaces. The $(S,T)$ channels in SNM up to 6-th order read 
\begin{eqnarray}
{\cal V}^{(0,0)} &=& \frac{3}{160}t^{(2)}_{2}(1-x^{(2)}_{2}) \rho k_F^2 +\frac{9}{560}t_{2}^{(4)}(1-x_{2}^{(4)})\rho k_{F}^{4}  \nonumber\\
        &+&\frac{1}{15}t_{2}^{(6)}(1-x_{2}^{(6)})\rho k_{F}^{6}\,, \label{eos:st2a} 
\end{eqnarray}
\begin{eqnarray}
{\cal V}^{(0,1)}&=&\frac{3}{16}t^{(0)}_{0}(1-x^{(0)}_{0})\rho + \frac{9}{160}t^{(2)}_{1}(1-x^{(2)}_{1})\rho k_F^2  \\
&+& \frac{27}{560}t_{1}^{(4)}(1-x_{1}^{(4)})\rho k_{F}^{4} +\frac{1}{5}  t_{1}^{(6)}(1-x_{1}^{(6)})\rho k_{F}^{6}\,, \label{eos:st2b} \nonumber
\end{eqnarray}
\begin{eqnarray}
{\cal V}^{(1,0)}&=&\frac{3}{16}t^{(0)}_{0}(1+x^{(0)}_{0})\rho + \frac{9}{160}t^{(2)}_{1}(1+x^{(2)}_{1})\rho k_F^2 \\
&+&\frac{27}{560}t_{1}^{(4)}(1+x_{1}^{(4)})\rho k_{F}^{4} +\frac{1}{5}  t_{1}^{(6)}(1+x_{1}^{(6)})\rho k_{F}^{6}\,, \label{eos:st2c}  \nonumber 
\end{eqnarray}
\begin{eqnarray}
{\cal V}^{(1,1)}&=&\frac{27}{160}t^{(2)}_{2}(1+x^{(2)}_{2})\rho k_F^2 +\frac{81}{560}t_{2}^{(4)}(1+x_{2}^{(4)})\rho k_{F}^{4}\nonumber\\    
         &+&\frac{3}{5}t_{2}^{(6)}(1+x_{2}^{(6)})\rho k_{F}^{6}\,.    \label{eos:st2d}     
\end{eqnarray}
Neither the tensor nor the spin-orbit terms contribute to the different $(S,T)$ channels. 
Using the expressions given in Eq.~(\ref{eos:st2a}-\ref{eos:st2d}), we have fitted the N2LO and N3LO parameters on the Brueckner-Hartree-Fock (BHF) calculations of Baldo \emph{et al.}~\cite{bal97}. The latter have been derived  from the microscopic Argonne $v14$ nucleon-nucleon two-body interaction plus the Urbana model for the three-body term. 
For completeness, we have also considered the chiral effective field ($\chi$-EFT) calculations at low k of Hebeler \emph{et al.}~\cite{heb11}, which cover a narrower density interval than BHF results.

We have found that no density-dependent term is needed to get a satisfactory fit. However, the resulting parameters give a too low value for the effective mass, $m^*/m \simeq 0.4$. We have thus included, on top of the N$\ell$LO pseudo-potential, a density-dependent term. 
Actually, the effective mass does depend only on the combinations $3 t_1^{(n)}+(5+4 x_2^{(n)}) t_2^{(n)}$, but the presence of a density-dependent term produces a readjustment of the other parameters $t_1^{(n)}$, thus modifying the value of $m^*/m$, an effect which has already been observed with the standard Skyrme interaction {\cite{gor03,kou12,klu09}. In general, such a density-dependent term should be obtained by applying the method of \cite{car08,rai11,dav13} to three-, four-, ... body interactions. However, for the sake of simplicity, we have considered the standard Skyrme effective density-dependent term $\frac{1}{6} t_3 (1+x_3 P_{\sigma}) \rho^{\alpha}$. Its contribution to all the formulae presented here can be immediately written, adding a similar contribution of $t_0^{(0)}$, with the replacements $t_0^{(0)} \to t_3 \rho^{\alpha}/6$ and $x_0^{(0)} \to x_3$. To keep a value of the effective mass $m^*/m \simeq 0.7$, we have fixed 
these parameters with two popular choices for $\alpha$, namely 1/6 and 1/3. 
It should be noticed that our conclusions do not depend on this particular choice. Furthermore, it is worth reminding that the present parameters should be considered only as a starting point for a more complete fit which should include finite nuclei observables. 

\begin{figure}[!h]
\begin{center}
\includegraphics[width=0.37\textwidth]{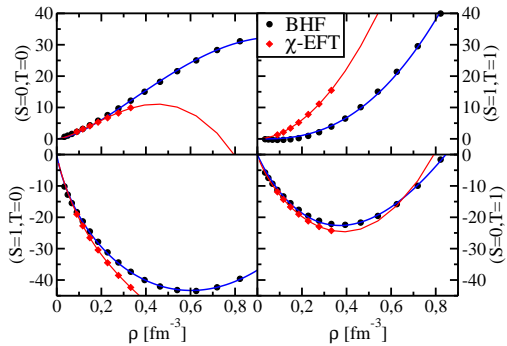}
\end{center}
\caption{N2LO fits of ${\cal V}^{(S,T)}(\rho)$ to the BHF (full dots) and $\chi$-EFT (full diamonds) results.}
\label{fig:diff}
\end{figure}

The N1LO interaction can fit the {\it ab-initio} these results only at low densities~\cite{les06}. On the contrary, a fair agreement is obtained with N2LO and N3LO. In Fig.~\ref{fig:diff} are displayed as a function of the density the results for the four $(S,T)$ channels as obtained at the N2LO order. Dots and diamonds are the BHF~\cite{bal97} and $\chi$-EFT~\cite{hebeler} results, respectively. Both are in reasonable agreement, except for the (1,1) channel. The N3LO fit of BHF channels stay on top of the N2LO, pointing towards a convergence of the fits in that density range of values. This is not the case with the $\chi$-EFT results, because the data at our disposal are limited to density values $\le 0.33$~fm$^{-3}$. The extrapolation to higher density values lead to very different N$\ell$LO, so that the fitted parameters are not reliable enough. In the following we thus only refer to the fit to BHF results.
In Tab.~\ref{tab:nall}, are given the numerical values of the central parameters of the pseudo-potential, here called VLyB6$\ell$ and VLyB3$\ell$, where $\ell$ refers to the N$\ell$LO order, and the density dependent choice is $\alpha=1/6, 1/3$, respectively. 


\begin{widetext}
\begin{table}[!h]
\begin{center}
\caption{The parameters of the VLyB6$\ell$ and VLyB3$\ell$ pseudo-potentials, fitted using a standard density dependent term. 
Missing entries are zero, and n.d. stands for not determined (see text).}
\begin{tabular}{cc|cc|cc||cc|cc}
\hline
\hline
 \multicolumn{2}{c|}{} & \multicolumn{4}{c||}{$\alpha=1/6$, $t_3$=13763 MeV fm$^{3(1+\alpha)}$, $x_3=1$} &   \multicolumn{4}{c}{$\alpha=1/3$, $t_3$=8000 MeV fm$^{3(1+\alpha)}$, $x_3=1$} \\
   \hline
 \multicolumn{2}{c|}{} & \multicolumn{2}{c|}{VLyB62} &   \multicolumn{2}{c||}{VLyB63}
 & \multicolumn{2}{c|}{VLyB32} &   \multicolumn{2}{c}{VLyB33}   \\
   \hline
$n$& i &$t^{(n)}_{i}$ [MeVfm$^{3+n}$] & $x^{(n)}_{i}$&$t^{(n)}_{i}$ [MeVfm$^{3+n}$] & $x^{(n)}_{i}$
&$t^{(n)}_{i}$ [MeVfm$^{3+n}$] & $x^{(n)}_{i}$&$t^{(n)}_{i}$ [MeVfm$^{3+n}$] & $x^{(n)}_{i}$ \\
0 &0& -2394.15 & 0.632433& -2500.49&0.598658       & -1491.46 & 0.409966& -1660.94&0.395793\\
 2&1&-19.381 &35.182 & 337.793&-2.18967                &133.037 &-3.97968 & 702.317&-0.534133\\
 2&2&513.2670&-1.01914& 587.22&-1.08109              &513.2670&-1.01914& 587.22&-1.08109\\
 4&1& 9.63577& 3.65615&-116.489 &-0.477614          & -14.6617& -0.745637&-215.685 &0.201958\\
 4&2 &-65.3664& -1.22006& -108.342&-1.33548         &-65.3664& -1.22006& -108.342&-1.33548\\
 6&1 &               &                 &2.84983 & -0.161801    &               &                 &4.54218 & 0.271071\\
 6&2 &               &                 &1.23644& -1.51102       &               &                 &1.23644& -1.51102\\
 \hline 
 \multicolumn{2}{c|}{} &  $t_{o}^{(n)}$ &  $t_{e}^{(n)}$&$t_{o}^{(n)}$ &  $t_{e}^{(n)}$ 
 &  $t_{o}^{(n)}$ &  $t_{e}^{(n)}$&$t_{o}^{(n)}$ &  $t_{e}^{(n)}$  \\
 2     &                         & 401.816 &         n.d. &  608.126     &  n.d.  & 401.816 &   n.d &  608.126     &  n.d.  \\
 4      &                        &-12.0604 & 294.851 & -72.78&-504.419      &-12.0604 & -225.429 & -72.78&-221.001\\
 6      &                        &                &           & 20.7379  &107.776     &                &           & 20.7379  &57.3209\\
 \hline
 & &\multicolumn{2}{c|}{$W_{0}$=245.741 [MeVfm$^{5}$]} & \multicolumn{2}{c||}{$W_{0}$=241.748 [MeVfm$^{5}$]}
 &\multicolumn{2}{c|}{$W_{0}$=245.741 [MeVfm$^{5}$]} & \multicolumn{2}{c}{$W_{0}$=241.748 [MeVfm$^{5}$]}  \\
\hline
\hline
\end{tabular}
\label{tab:nall}
\end{center}
\end{table}
\end{widetext}

Let us comment now on the fitted central parameters. 
A first trend is that within a given N$\ell$LO pseudo-potential, the $t_i^{(n+1)}$ parameters are always smaller, in absolute value, than $t_i^{(n)}$, which is a good signal about the convergence of the expansion in gradients. 
However, the value of a given parameter can be very different, even in sign, when going from N2LO to N3LO since the asymptotic behavior ({\it i.e.} the maximum power in the fit), is no longer governed by the same parameters. Thus, $t_1^{(4)}$ in the N2LO fit, is constrained by the asymptotic behavior when we deal with N2LO, but it can take a very different value when higher order terms are considered in N3LO. 
By inspecting Eqs~(\ref{eos:st2a})-(\ref{eos:st2b}) one can see that the density dependent term only enters into the fit of channels (0,1) and (1,0), and consequently the space-odd parameters $t_2^{(n)}, x_2^{(n)}$ are independent of the choice of $\alpha, t_3, x_3$. The N3LO space-even parameters are of the same sign and order of magnitude for the two selected density dependent terms. 
Using the central parameters given in Tab.~\ref{tab:nall}, we have calculated some basic SNM properties at the saturation density $\rho_{\rm sat}$. With VLyB62 interaction, we obtain the following values: $\rho_{\rm sat}= 0.179$~fm$^{-3}$, $E/A=-15.50$~MeV, $m^{*}/m=0.789$, and $K_{\infty}=220.8$~MeV. With VLyB63 these values are 0.169~fm$^{-3}$, $-15.72$~MeV, 0.762, and 201.9~MeV, respectively. 
The interactions VLyB32 and VLyB33 give comparable results.  

To determine the spin-orbit and tensor contributions, we have to go one step further, and project the potential energy terms onto the $(J,L,S,T)$ sub-spaces, where $J$ and $L$ are the total and orbital angular momenta, respectively. We have used the partial waves provided by the BHF  microscopic calculations~\cite{bal14}. In Ref.~\cite{dav14k} we have given explicit expressions for the potential energy projections ${\cal V}(^{2S+1}L_J)$, where we use the standard spectroscopic notation. 
The previous $(S,T)$-projections are obtained by summing up all these partial waves, and are independent of the spin-orbit and tensor parameters. This is also the case for the spin-singlet partial waves, which are thus fixed from the fitted central parameters. 
On the contrary, triplet spin partial waves do depend on the spin-orbit and tensor terms, leading to the removal of the degeneracy between the different partial waves for given $J$. 

The triplet partial waves relevant for our discussion are the following
\begin{eqnarray}
\label{ondeS}
{\cal V}(^{3}S_1)&=&\frac{3}{16}t^{(0)}_{0}(1+x^{(0)}_{0})\rho
+ \frac{9}{160}t^{(2)}_{1}(1+x^{(2)}_{1})\rho k_{F}^{2} \\
&+&\frac{9}{280}t_{1}^{(4)}(1+x_{1}^{(4)})\rho k_{F}^{4} 
+\frac{1}{10}t_{1}^{(6)}(1+x_{1}^{(6)})\rho k_{F}^{6} \;,\nonumber 
\end{eqnarray}
\begin{eqnarray}
{\cal V}(^{3}P_0)&=&\frac{3}{160}t^{(2)}_{2}(1+x^{(2)}_{2})\rho k_{F}^{2}+\frac{9}{560}t_{2}^{(4)}(1+x_{2}^{(4)})\rho k_{F}^{4}  \nonumber\\
&+&\frac{3}{50}t_{2}^{(6)}(1+x_{2}^{(6)})\rho k_{F}^{6}  +\frac{1}{40} W_0 \rho k_{F}^{2} \nonumber\\
&-& \frac{3}{80}t^{(2)}_{o}\rho k_{F}^{2} 
- \frac{9}{140}t_{o}^{(4)}\rho k_{F}^{4}-\frac{17}{750}t_{o}^{(6)}\rho k_{F}^{6}  \;,
\end{eqnarray}
\begin{eqnarray}
{\cal V}(^{3}P_1)&=&\frac{9}{160}t^{(2)}_{2}(1+x^{(2)}_{2})\rho k_{F}^{2}+\frac{27}{560}t_{2}^{(4)}(1+x_{2}^{(4)})\rho k_{F}^{4}  \nonumber\\
&+&\frac{9}{50}t_{2}^{(6)}(1+x_{2}^{(6)})\rho k_{F}^{6} +\frac{3}{80} W_0 \rho k_{F}^{2} \nonumber\\
&+& \frac{9}{160}t^{(2)}_{o}\rho k_{F}^{2} 
+ \frac{27}{280}t_{o}^{(4)}\rho k_{F}^{4}+\frac{17}{500}t_{o}^{(6)}\rho k_{F}^{6}  \;,
\end{eqnarray}
\begin{eqnarray}
{\cal V}(^{3}P_2)&=&\frac{3}{32}t^{(2)}_{2}(1+x^{(2)}_{2})\rho k_{F}^{2}+\frac{9}{112}t_{2}^{(4)}(1+x_{2}^{(4)})\rho k_{F}^{4}  \nonumber\\
&+&\frac{3}{10}t_{2}^{(6)}(1+x_{2}^{(6)})\rho k_{F}^{6} - \frac{1}{16} W_0 \rho k_{F}^{2} \nonumber\\
&-& \frac{3}{160}t^{(2)}_{o}\rho k_{F}^{2} 
- \frac{9}{280}t_{o}^{(4)}\rho k_{F}^{4}-\frac{17}{1500}t_{o}^{(6)}\rho k_{F}^{6}  \;,
\end{eqnarray}
\begin{eqnarray}
{\cal V}(^{3}D_1)&=&\frac{9}{2800}t_{1}^{(4)}(1+x_{1}^{(4)})\rho k_{F}^{4}+\frac{1}{50}t_{1}^{(6)}(1+x_{1}^{(6)})\rho k_{F}^{6}  \nonumber\\
&-& \frac{9}{1000}t_{e}^{(4)}\rho k_{F}^{4} -\frac{7}{1500}  t_{e}^{(6)} \rho k_{F}^{6}\;,\\
%
\label{ondeF}
{\cal V}(^{3}F_2)&=&\frac{1}{70}t_{2}^{(6)}(1+x_{2}^{(6)})\rho k_{F}^{6}  -\frac{3}{875}t_{o}^{(6)}\rho k_{F}^{6} \,.
\end{eqnarray}
It is worth noticing that they do not depend on the tensor parameter $t_e^{(2)}$. Indeed, for the equation of state, only diagonal elements appear and the $t_e^{(2)}$ term connects only states with $L=0$ with states with $L=2$. Consequently, $t_e^{(2)}$ cannot be determined with this type of calculation. 

By keeping the central parameters previously fitted to the $(S,T)$-channels, it is possible to determine the tensor and spin-orbit parameters $t_{o}^{(2,4,6)},t_{e}^{(4,6)},W_{0}$ by fitting directly to the BHF partial waves. However, the BHF calculation we use for the fit provides in fact two mixtures, namely $^3S_1+ ^3D_1$, and $^3P_2+ ^3F_2$. We have decided to fit the tensor parameters $t_e^{(4)}, t_e^{(6)}$ to the former of these combinations, and the spin-orbit $W_0$ and the tensor $t_o^{(n)}$ parameters to the partial waves $^3P_0$ and $^3P_1$. The comparison with the remaining BHF partial waves is in fact a prediction. In Fig.~\ref{fig:jlst} are displayed the comparisons obtained at the N3LO level. We don't display the N2LO results, because in contrast to the $(S,T)$ channels, the fits are not so good, with notorious differences, in particular for the $^1S_0$ and $^3S_1+^3D_1$ waves. 

\begin{figure}[h]
\begin{center}
\includegraphics[width=0.4\textwidth]{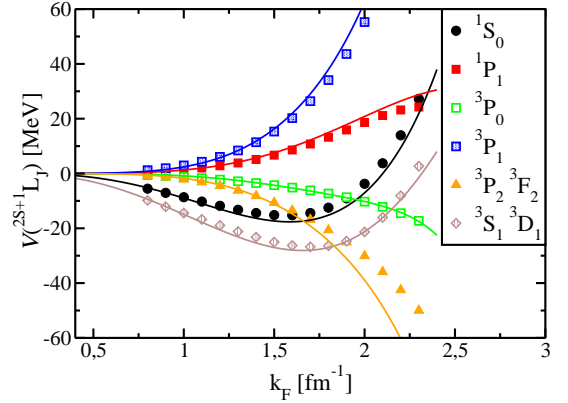}\\
\end{center}
\caption{(Color online) Comparing the partial waves ${\cal V}(^{2S+1}L_J)$. The dots are the BHF results, and the lines are the results of our fit with VLyB3.}
\label{fig:jlst}
\end{figure}

One can see that the N3LO results are quite satisfactory, with the only appreciable difference for the $^3P_2+^3F_2$ coupled
waves at high density values. Since the $F$ wave receives contribution of order $ \rho k_F^6$ only (see Eq. \ref{ondeF}), it is clear that this discrepancy reflects a missing contribution of higher order (N4LO), as expected for a partial wave with $\ell =3$.

Analogously to the central $t_2^{(n)}, x_2^{(n)}$ parameters, the spin-orbit $W_0$ and tensor $t_o^{(n)}$ parameters are fixed whatever the choice made for the density-dependent term, as can be explicitly seen by inspecting Eqs~(\ref{ondeS})-(\ref{ondeF}). The fits give practically the same $W_0$ value (about twice the typical standard Skyrme values), with a positive sign. The same is true for $t_o^{(2)}$, whose N2LO and N3LO values are of the same order of magnitude, being positive in both cases. Since this result is also in agreement with our previous work on Landau parameters \cite{dav14c}, we have now consistent evidence that the parameter $t_o^{(2)}$ has to be restricted to positive values. Finally, one can also see that  
within a given N$\ell$LO pseudo-potential, the $t_{o,e}^{(n+1)}$ parameters are always smaller, in absolute value, than 
$t_{o,e}^{(n)}$, pointing again to a good convergence of the expansion in gradients.

\emph{Conclusions.} The inclusion of higher order derivative terms to the Skyrme pseudo-potential is of fundamental importance to improve the spectroscopic qualities of the model, without loosing the previous expertise gained by the nuclear structure community since its very first applications in the early 70s.
In this letter, we have demonstrated that the extra derivative terms can heal some major problems of \emph{standard} Skyrme functionals. Performing a partial wave decomposition of the Skyrme pseudo-potential, we have been able to identify for the very first time, the contribution of the Skyrme tensor parameters to the $(S,T)$-channels and partial waves. Moreover, by fitting the BHF results, we have been able to derive a constraint on the sign of spin-orbit and tensor parameter $t_o^{(2)}$ in total agreement with our previous work on Landau parameters. With the fitted parameters, we have calculated, as a function of the density, quantities as Landau parameters, and the EoS of polarized matter, asymmetric matter, neutron matter and polarized neutron matter. No instabilities appear in the range of densities $\le 0.3$~fm$^{-3}$ which is relevant for finite nuclei.
This set of parameters can thus be seen as a good starting point for a fit protocol which also includes constraints from finite nuclei.

Since our analysis is mainly based on BHF calculations, the question arises as to wether our method is of general reliability. We have seen that BHF and $\chi$-EFT results are in fair agreement at low densities, except in the (1,1)-channel. As $\chi$-EFT results are not yet available at high density values, we have not access to the asymptotic behavior and a detailed comparison of the fits is not yet possible. However, 
our method enough to expect good fits to other {\it ab-initio} results, offering thus the possibility of comparing among them, getting in touch with finite nuclei results through an effective $N\ell LO$ pseudo-potential.

\emph{Acknowledgments} We thank M. Baldo and K. Hebeler for providing us with  his BHF results.
 The work of JN has been supported by grant FIS2011-28617-C02-2, Mineco (Spain). A.P. acknowledges the University of Jyv\'askyl\"a and the FIDIPRO group for hospitality during which part of this work has been done.

\bibliography{biblio}


\end{document}